# Heating Isotopically Labelled Bernal Stacked Graphene: a Raman Spectroscopy Study

Johan Ek Weis, Sara Costa, Otakar Frank and Martin Kalbac*

*J. Heyrovský Institute of Physical Chemistry, Academy of Sciences of the Czech Republic, v.v.i., Dolejškova 3, CZ-18223 Prague 8, Czech Republic.*
*Email: martin.kalbac@jh-inst.cas.cz

**Abstract**

One of the greatest issues of nanoelectronics today is how to control the heating of the components. Graphene is a promising material in this area and it is essential to study its thermal properties. Here, the effect of heating a bilayer structure was investigated using *in situ* Raman spectroscopy. In order to observe the effects on each individual layer, an isotopically labelled bilayer graphene was synthesized where the two layers are composed of different carbon isotopes. Therefore, the frequency of the phonons in the Raman spectra is shifted in relation to each other. This technique was used to investigate the influence of different stacking order. It was found that in bilayer graphene grown by chemical vapor deposition (CVD) the two layers behave very similarly, for both Bernal stacking and randomly oriented structures, while for transferred samples the layers act more independently. This highlights a significant dependence on sample preparation procedure.

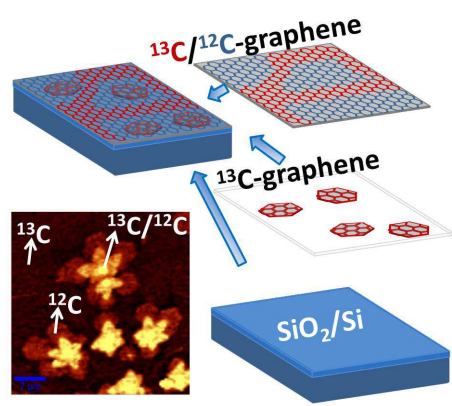

Bernal (AB) stacked bilayers have been considered for most applications.[1] Now turbostratic bilayer graphene is getting more and more interest and new applications are emerging, such as bilayer pseudospin field-effect transistors (BISFET).[2] However, one of the greatest issues of nanoelectronics today is how to control the heating of the components.[3] It is therefore essential to study the thermal properties of any new electronic components. The temperature can potentially be used to tune the properties of new devices.

Raman spectroscopy is a powerful tool to investigate carbon nanomaterials. It is a non-destructive technique and can yield a lot of data on the structure of the material. In the case of graphene, information about strain, disorder, number of layers, doping and other properties can be obtained.[4] A typical defect-free graphene sample shows a Raman spectrum with the G band, at around 1585 cm$^{-1}$, and the 2D (or G') band at around 2700 cm$^{-1}$ (for an excitation energy of 2.54 eV). Additional features can be observed if graphene containing defects, e. g. D and D' bands at ~1350 and 1620 cm$^{-1}$, respectively (at 2.54 eV).

Besides the perturbations mentioned above, the frequency of the Raman features is also very sensitive to temperature,[5] and to atomic mass. The first allows the study of the effects of heating and yields information about the thermal transport.[6] Heating experiments on monolayer,[7] and multilayer[8] graphene have been reported, including the transition between graphene and graphite. Several experimental studies used the laser as the heating source and the position of the Raman features as guidelines to determine the local temperature rise.[9-11] The heating occurs locally, and there is no heating equilibrium, as only a small area of the material is under the laser spot, and gets heated. Therefore, the Raman features change in a different way than in processes carried out in heat cells. Besides that, this kind of study is usually carried out in air and the effects on the Raman spectra comprise the presence of oxygen and other molecules. On the other hand, studies carried out in heating cells and under gas flow avoid the presence of external molecules and focus more on the intrinsic properties of graphene.[12]

Studying bilayer graphene can also be challenging because even though the shape and width of the 2D band can be used to determine the number of layers in graphene, it can still be difficult to study individual layers in a multilayer structure since they yield very similar signals. Here, this problem is resolved by studying a bilayer where the two layers are composed of different isotopes of carbon.[13] When the atoms in the two layers have different mass, the frequency of the phonons in the Raman spectra are shifted with respect to each other and information about the individual layers can be obtained. A similar structure was investigated in reference [12], where the bilayer was created by subsequent transfer of single layers of graphene whereas the bilayer in the presented study is grown by CVD in a single process. The technique of subsequent transfer of individual layers has also been used to create and study a three layered graphene structure, with one layer consisting of $^{12}$C, one of $^{13}$C and the middle layer of a mixture of the two isotopes.[14] The individual layers could clearly be identified in the Raman spectra due to the different mass.

Whether or not the type of substrate influences the Raman signal of monolayer graphene has been investigated.[15,16] It was found that for graphene fabricated by micromechanical cleavage of graphite, the influence of the underlying substrate is negligible, whereas a shift was found for samples grown on SiC [16]. However, a G peak shift of 5 cm$^{-1}$ was found for graphene on a sapphire substrate.[15] The interaction between the graphene and the substrate is important during heating experiments. It was shown previously, [12] that heating of the structure induces stress due to different coefficient of thermal expansion between the graphene and the SiO$_2$/Si substrate. Importantly, it was found that this stress strongly influences the bottom layer, which is in contact with the substrate, whereas the top layer remained almost unaffected. A similar effect was found on partly suspended monolayer,[17] where the G band frequency of the supported part was shifted, whereas the suspended part remained stable during a heating cycle.

The individual layers in a bilayer structure can either be ordered in a Bernal stacking or randomly oriented (turbostratic). In this work, we compare how these different structures behave during heating and also use isotope labelling to simultaneously investigate the top and bottom layers separately. The difference between a bilayer grown by CVD in one step is also compared with a bilayer fabricated by subsequent transfer of two individual monolayers of graphene, as studied in [12].

The graphene sample investigated here was synthesized by chemical vapor deposition (CVD). The carbon feedstock was switched from $^{12}$CH$_4$ (methane), to $^{13}$CH$_4$ during the growth. Consequently, graphene islands constituting atoms of carbon 12 ($^{12}$C) were first grown with small adlayers at the nucleation points. With the introduction of $^{13}$CH$_4$, the graphene continued to grow, but now containing carbon 13 ($^{13}$C) atoms. The adlayers also expanded, but at a slower rate, forming a bilayer of $^{12}$C/$^{13}$C. Bilayer graphene composed of only one isotope can also be found, in small areas around the nucleation points. Both turbostratic (T) and Bernal (AB) stacked bilayer graphene grains can be found within one sample. The investigated bilayers have therefore undergone the same treatment regardless of stacking order. The material was subsequently transferred to a SiO$_2$/Si substrate. A summary of these steps leading to an isotope labelled bilayer graphene structure on a SiO$_2$/Si substrate is shown in figure 1.

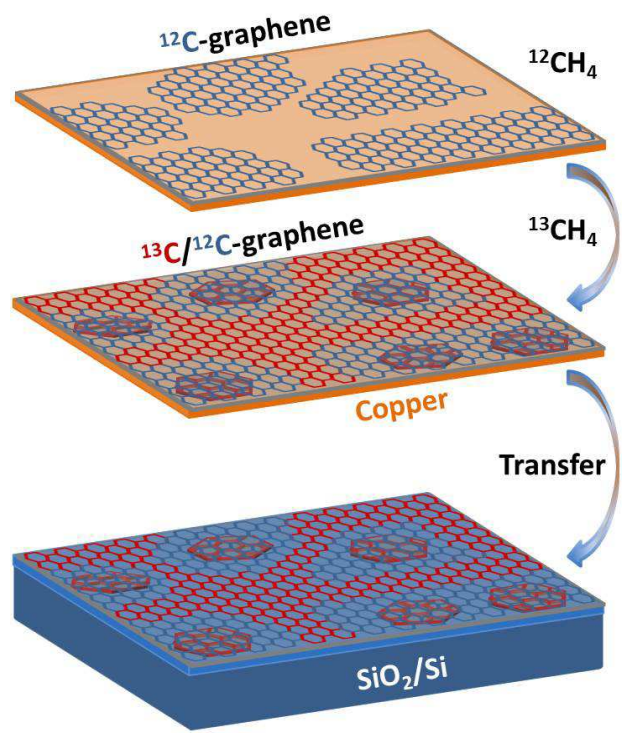

**Figure 1- Schematic view of the isotopically labelled 2-LG growth by CVD. Firstly, $^{12}$C graphene is grown around nucleation points. Secondly, $^{12}$CH$_4$ is replaced by $^{13}$CH$_4$ and graphene islands containing $^{13}$C atoms are formed. The introduction of the second isotope also allows the $^{12}$C grains to keep growing. The material was then transferred onto a SiO$_2$/Si substrate.**

Figure 2 shows typical Raman spectra of bilayer graphene comprised of $^{12}$C and $^{13}$C isotopes. The downshift of the frequency of the bands in the $^{13}$C sample originates from its higher mass. The shift can be calculated by Eq 1:

$$\frac{\omega_{12} - \omega_{13}}{\omega_{12}} = 1 - \left[\frac{12 + c_0}{12 + c_{13}}\right]^{1/2}$$

where $\omega_{12}$ and $\omega_{13}$ are the frequencies of the Raman peak in a $^{12}$C and $^{13}$C sample, respectively, whereas $c_{13}=0.99$ and $c_0=0.0107$ are the concentrations of $^{13}$C in the enriched sample and in the $^{12}$C sample, respectively. The value of $c_0$ is thus the natural abundance of $^{13}$C (1.07 %), which is expected to be found in the $^{12}$C sample, whereas the value of $c_{13}$ comes from the purity of the $^{13}$C precursor (99%). At room temperature, the expected downshifts of the G and 2D bands according to equation 1 are 56 and 95 cm$^{-1}$, respectively, which is in good agreement with our experimental results.

Raman spectroscopy was performed with a laser excitation energy of 2.54 eV (488 nm). Figure 2a shows the Raman spectra of different types of bilayer graphene obtained from different regions in the sample: Bernal stacked 2-LG (1), and turbostratic 2-LG (2), $^{12}$C 2-LG (3), and $^{13}$C 2-LG (4). Figures 2b and 2c show Raman maps of the total intensity of 2D band for AB and T stacked graphene grains. Several differences are noteworthy when observing these spectra. 2-LG containing only $^{13}$C shows a G band at around 1530 cm$^{-1}$, while for the one containing only $^{12}$C, the G band is at around 1585 cm$^{-1}$, which is in agreement with equation 1. In the spectra labelled with both isotopes, the two peaks can be observed simultaneously. A similar behavior is observed for the 2D band of turbostratic 2-LG, where two peaks can also be observed (at ~2615 and 2700 cm$^{-1}$ for $^{13}$C and $^{12}$C, respectively). For the AB stacked 2-LG (both CVD and exfoliated), the line shape of 2D band changes and the isotope modes are no longer well separated. This asymmetric line shape is typically assigned to AB stacked 2-LG and it comprises four peaks that arise from the different electron-phonon scattering processes, as described in [18]. This also leads to an increase in the full width at half maximum (FWHM), from ~40 cm$^{-1}$ for T 2-LG to 50 cm$^{-1}$ for exfoliated AB 2-LG, and 106 cm$^{-1}$ for labelled CVD AB 2-LG. This difference was previously reported by Fang *et al.*, and can be explained by the influence of the mass difference on some of the vibrational modes.[19] Other less intensive bands, such as D+D'' (2450 cm$^{-1}$) were also observed.

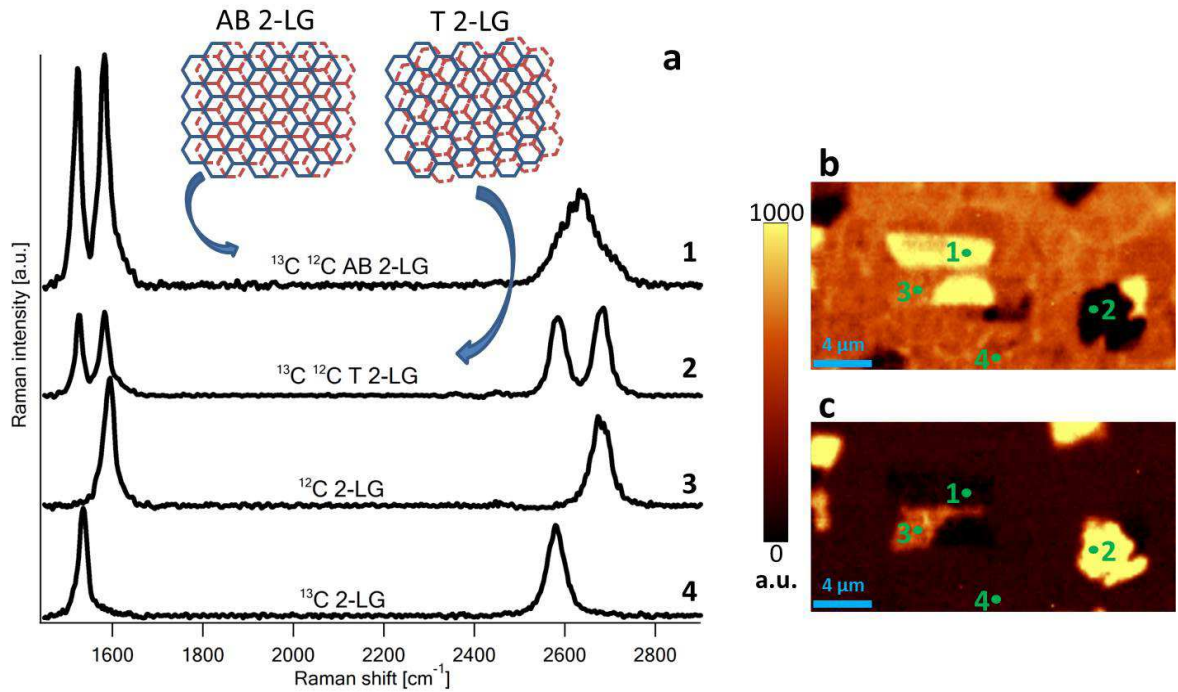

**Figure 2-** (a) Raman spectra of bilayer graphene: $^{13}C/^{12}C$ with Bernal stacking, AB 2-LG (1), and $^{13}C/^{12}C$ turbostratic, T 2-LG (2), monoisotopic $^{12}C$ 2-LG (3) and $^{13}C$ 2-LG (4). The spectra are offset for clarity. (b), (c) Raman maps of the total intensity of 2D band of different graphene grains. In (b) AB stacked areas are highlighted in bright colors, whereas in (c) the image shows turbostratic areas. All the spectra were acquired at room temperature. The scale bar is 4 µm.

A clear difference is observed between the 2D band shape of Bernal stacked and turbostratic grains. For the latter, the peaks for the $^{12}C$ and $^{13}C$ layers in the 2D band are well separated (see Figure 2), and can be fitted with single Lorentzians, whereas for AB 2-LG the fitting process is more complex. Fang et al. [19], and Araujo et al. [20] both suggested the use of 8 Lorentzian peaks, comprising the effects of the $^{13}C$ and $^{12}C$ presence. If each layer is composed of different isotopes, eight processes can be rationalized directly, four assigned to $^{12}C$ and four assigned to $^{13}C$, with frequencies shifted from each other according to equation 1. However, processes involving both isotopes should also be taken into account due to the strong coupling in AB 2-LG, extending the number of peaks in the 2D band to 12. In order to clarify the two-phonon processes in labelled AB stacked samples further studies are currently being carried out, but will not be discussed here as it is not the goal of this work.

The Raman features described above are sensitive to the intrinsic properties of the material, and also to extrinsic factors, such as doping. In order to eliminate impurities and possible dopants, the sample was first annealed by heating it up to 1173 K (cycle 1). This removes oxygen and any other impurities from the surface of the graphene. A hysteresis effect is observed during this first heating cycle. That is, the peak positions during the heating and cooling stages are shifted. This is due to the impurities, which dope the graphene and thereby shift the Raman peaks from their intrinsic positions. However, no such hysteresis effect is observed in the second heating cycle, indicating that the impurities are removed and that the intrinsic behavior of the bilayer graphene is measured. A similar behavior was reported previously.[7,12,21] This behavior with hysteresis in the first cycle but not in subsequent ones is reproduced in all experiments performed.

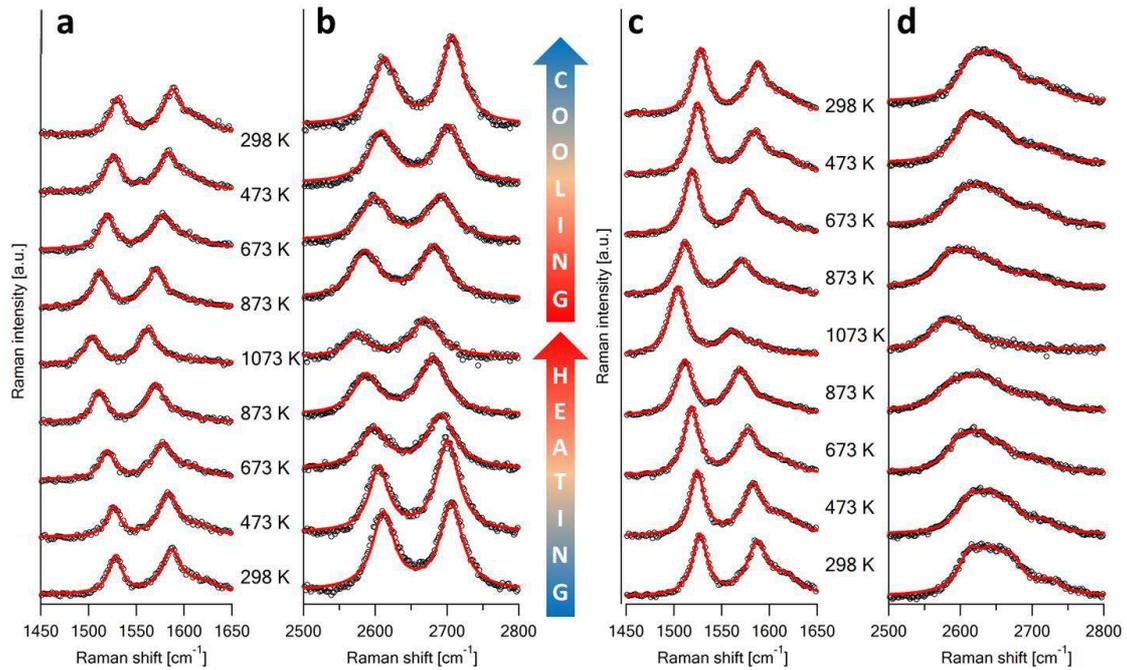

**Figure 3- In situ Raman spectra of T 2-LG (a and b), and AB 2-LG (c and d), at different temperatures, using excitation energy of 2.54 eV. The points represent the experimental data, and the solid lines are the convoluted Lorentzian line shapes used to fit the data. The spectra are offset for clarity.**

Figure 3 shows the Raman spectra during the second heating/cooling cycle for a turbostratic (a, b), and Bernal stacked (c, d) 2-LG, as can be seen from the 2D band shape. The frequency of the modes is well separated for the $^{12}$C and $^{13}$C layers and it is sensitive to the temperature, as can be seen from the downshifts as the material is heated. An upshift of all bands occurs when the sample is cooled, and the initial frequencies are recovered when room temperature is reached.

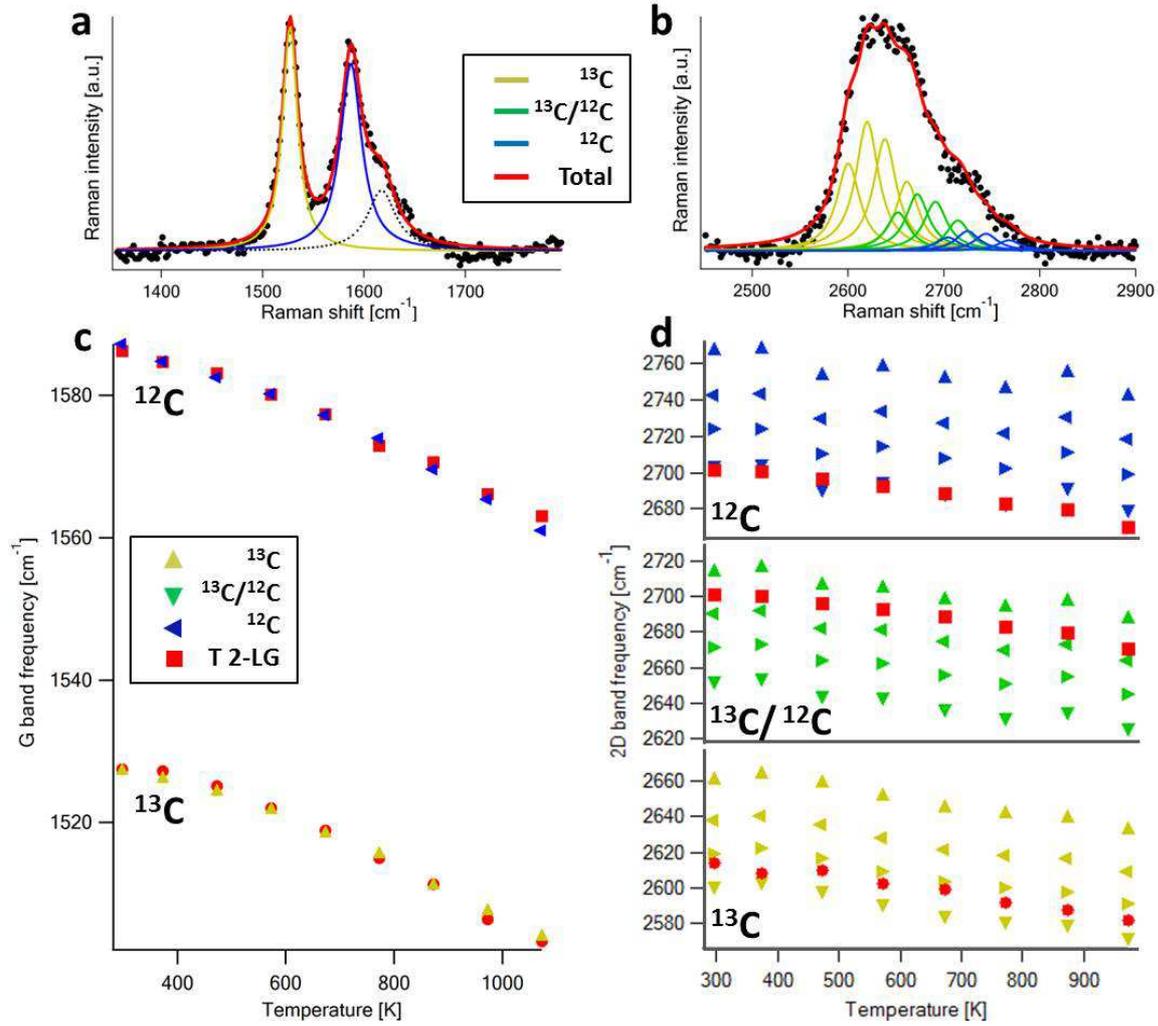

**Figure 4** – (a) The G band, and (b) the 2D band fits for AB 2-LG, using Lorentzian line shapes. The temperature dependence of (c) G band, and (d) 2D band. The red symbols represent T 2-LG, and blue, green, and yellow symbols are assigned to individual components in AB 2-LG of $^{12}$C, $^{13}$C/$^{12}$C, and $^{13}$C, respectively.

For a more clear observation, the frequency of the bands was plotted against the temperature (Figure 4). The G band for the T and AB bilayers shift very similarly during the heating cycle, see Figure 4c and 4d. The G band originating from the $^{12}$C of the turbostratic sample shifts at a rate of -0.033 ±0.001 cm$^{-1}$K$^{-1}$, whereas the $^{13}$C band shift is -0.035 ±0.002 cm$^{-1}$K$^{-1}$. However, in the AB stacked sample, there is no difference between the shifts of the G bands originating from $^{12}$C or $^{13}$C, both of them shift at a rate of -0.032 ±0.001 cm$^{-1}$K$^{-1}$. This can be compared to 0.026±0.001 cm$^{-1}$K$^{-1}$ and 0.044±0.001 cm$^{-1}$K$^{-1}$ shift rates for the top and bottom layer, respectively, which were found for a transferred bilayer structure in [12]. As all the extrinsic effects are expected to be eliminated during the first heating cycle, the difference between the slopes in [12] were attributed to the different interaction of the top and bottom layer with the substrate. In the CVD grown bilayer graphene studied here, the difference between the slopes is significantly smaller. However, a stronger coupling between the layers is expected for a CVD grown graphene multilayer, in comparison to a transferred structure. This explains the small difference found between the top and bottom layers here. The coupling between the layers in the AB stacked sample is expected to be stronger than in the turbostratic, but the difference is negligible. This can also be seen here as there is no difference between the shifts of the layers in the AB stacked sample, whereas the bottom $^{13}$C layer shifts more than the top $^{12}$C layer in the turbostratic samples.

There are two separated peaks in the 2D region for a turbostratic bilayer. One originates from the $^{12}$C and the other from the $^{13}$C layer. These shift very similarly as the temperature increases, -0.045 ±0.003 cm$^{-1}$K$^{-1}$ and -0.047 ±0.003 cm$^{-1}$K$^{-1}$ for the $^{12}$C and $^{13}$C, respectively. This can be compared to the transferred bilayer structure investigated in [12] with shift rates of -0.043 ±0.002 cm$^{-1}$K$^{-1}$ and -0.075 ±0.001 cm$^{-1}$K$^{-1}$ for the top and bottom layers, respectively. The lower shift of the top layer was associated with the intrinsic shift of the graphene layer, whereas the shift of the bottom layer was increased due to strain caused by interactions with the substrate. The turbostratic bilayer graphene studied here was grown in one step, rather than fabricated but two subsequent transfers, thus it behaves as a single unit, with both layers showing only intrinsic shifts.

The case of 2D band in the AB stacked 2-LG is more complex. As referred above, the 2D band can be fit by 12 peaks, as shown in figure 4c. The four peaks with the lowest frequency can be assigned to the $^{13}$C (yellow lines), the four peaks with highest frequency to the $^{12}$C (blue lines), and the four middle peaks involve both types of isotopes (green lines). Figure 4d shows the frequency shift of each 2D peak plotted with the temperature. Similarly to the G band, the frequency also downshifts with the increase of temperature. The peaks related to the $^{13}$C all shift at the same rate as for the turbostratic samples, -0.047 ±0.004 cm$^{-1}$K$^{-1}$, while the mixed peaks shift slightly less, -0.042 ±0.005 cm$^{-1}$K$^{-1}$. These shifts are still within the intrinsic limit, and as observed in the G band, both layers seem to shift similarly. This behavior is expected because of the stronger coupling between two layers with Bernal stacking. Unfortunately, the intensity of the $^{12}$C peaks is too low to obtain a reliable fit, due to the asymmetric shape of the 2D band.

Table 1- Least-squares fitted slopes for the G and 2D band shifts with temperature in $^{13}$C/$^{12}$C bilayer graphene, during heating and cooling half cycles.

|  | cm$^{-1}$K$^{-1}$ | | | | |
| --- | --- | --- | --- | --- | --- |
|  | $^{13}$C-G | $^{12}$C-G | $^{13}$C-2D | $^{13}$C/$^{12}$C -2D | $^{12}$C-2D |
| T heat | 0.035 ± 0.002 | 0.033 ± 0.001 | 0.047 ±0.003 | - | 0.045 ±0.003 |
| T cool | 0.035 ± 0.001 | 0.033 ± 0.002 | 0.045 ±0.003 | - | 0.044 ±0.003 |
| AB heat | 0.032 ± 0.001 | 0.032 ± 0.001 | 0.047 ±0.004 | 0.042 ±0.005 | - |
| AB cool | 0.032 ± 0.001 | 0.031 ± 0.002 | 0.046 ±0.004 | 0.043 ±0.004 | - |

Table 1 presents the shift rates of the G and 2D bands for all the samples, during the whole heating/cooling cycle. In general, the bands shift at similar rates during heating and cooling, recovering their intrinsic frequencies after the cycle is complete.

It has been found that in a weakly coupled bilayer, the top layer contains significantly more wrinkles than the bottom layer and that almost all edges of the top layer are folded [13]. It was discussed that the wrinkles formed to release any local stress of the top layer and were allowed to form due to the weak interaction between the graphene layers. The fact that no significant difference in amount of wrinkles is found between the layers here and that no folded edges are observed are further indications that the layers in the CVD grown bilayer are strongly coupled and behave like one unit, which is in agreement with the findings of reference [13]. The interlayer coupling has been estimated for AB stack layers (~0.4 eV).[22] However, for turbostratic layers this value depends on the rotation angle and on the relation between the position of the atoms in the two layers. Turbostratic layers can therefore either behave as two individual layers or similarly to AB stacked layers depending on the orientation.[23] In the case of CVD grown turbostratic graphene, the rotation angles are close to AB stacked grains, in a way that the two kinds behave similarly.

In summary, CVD grown isotope labelled bilayer graphene was studied during heating cycles from room temperature up to 1173 K. The isotope labelling allowed us to follow the heating effect simultaneously and independently in the top and bottom layers. The behavior of both turbostratic and AB stacked bilayers were investigated. It was found that the coupling between the two layers is strong for both stacking orders, despite the fact that turbostratic grains still show separated signals for the two layers, with intensities similar to monolayer graphene. In fact, a small difference can be observed between the layers in turbostratic bilayers and AB stacked samples during heating. These findings differ from the ones obtained for bilayers constructed by subsequent transfer of two single layers of graphene, where the layers were found to be decoupled with the top layer floating freely on top of the bottom layer. This difference highlights the fact that the two layers grown by CVD are strongly coupled. We, therefore, show that the behavior of a bilayer structure is strongly influenced by the synthesis method.

**Experimental methods**

*Growth*: The graphene samples were synthesized by chemical vapor deposition (CVD). A copper foil was heated to 1273 K and annealed for 20 min under a flow of 50 sccm $H_2$. 1 sccm of methane ($^{12}CH_4$) was then introduced to the chamber for 90 s, which resulted in the growth of graphene constituting $^{12}C$. The $^{12}CH_4$ was turned off and a flow of 1 sccm $^{13}CH_4$ was introduced for 30 min. The sample was cooled down to room temperature.

*Transfer:* The graphene was subsequently transferred to a $SiO_2$/Si substrate, as reported previously.[24] In brief, the substrate was covered by poly(methyl methacrylate) (PMMA). The copper was subsequently etched away and the PMMA covered graphene was transferred to a $SiO_2$/Si substrate. Residual PMMA was removed by heating the sample to 773 K in a hydrogen and argon atmosphere.

*Thermal treatment:* The sample was placed in a heating cell (Linkam), and a flow of He gas was supplied in order to keep the chamber free from oxygen. The heating process was carried out in two heating/cooling cycles, from room temperature to 1173 K.

*Characterization: In situ* Raman spectra were acquired every 200 K and every 100 K, for cycle 1 and cycle 2, respectively. Raman measurements were performed using a LabRAM HR Raman spectrometer (Horiba Jobin-Yvon) and an Ar/Kr laser (488 nm, Coherent). A 50x objective was used, providing a laser spot of about 1 μm. The laser power was 1 mW outside of the heating chamber, but about 0.7 mW at the sample after going through a sapphire window and an aperture in a heat radiation shield. The Raman maps were acquired using a WiTec Alpha300, with a 532 nm laser excitation, and 100x objective.

**Acknowledgments**

The authors Johan Ek Weis and Sara Costa contributed equally to this manuscript. The authors acknowledge the support of MSMT ERC-CZ project (LL 1301).